# Resonant mode flopping in modulated waveguiding structures

Yaroslav V. Kartashov,[1] Victor A. Vysloukh,[2] and Lluis Torner[1]

[1]ICFO-Institut de Ciencies Fotoniques, Mediterranean Technology Park, and Universitat Politecnica de Catalunya, 08860 Castelldefels (Barcelona), Spain

[2]Departamento de Fisica y Matematicas, Universidad de las Americas – Puebla, Santa Catarina Martir, 72820, Puebla, Mexico

We put forward the concept of resonant, Rabi-like flopping and adiabatic transitions between confined light modes in properly modulated multimode waveguides. The phenomenon is shown to take place in both, the linear and the nonlinear regimes. In addition, we find that the mode transitions occur not only in simple geometries, but also in complex confining multimode structures. The phenomenon is analogous to the familiar stimulated state transitions that occur in multilevel quantum systems.

More than seventy years ago I. Rabi published his seminal paper about two-level quantum system resonantly interacting with an electromagnetic wave [1]. The phenomenon is universal and occurs in a variety of physical settings: Atomic energy levels; the valence and conduction bands in semiconductors; spin orientations in a magnetic field; etc. Rabi found that when the electromagnetic wave is turned on the population of the higher energy levels grows. Once the stimulated transition is complete, the electromagnetic wave stimulates the emission of radiation so that the system returns to its ground state. This revival process is periodic and its frequency (Rabi flopping frequency) increases with amplitude of the electromagnetic wave. Since Rabi's seminal paper there has been a huge progress in studies of quantum revivals in atomic and molecular systems, as well as of related revival phenomena in condensed matter and optical systems (for a recent review see [2]).

In classical Optics, revivals appear also, e.g., as periodical self-imaging or Talbot effects. This effect has been observed in a variety of physical systems (see [3-6] for a few examples). Introducing longitudinal modulations in optical structures can lead to similar phenomena. Thus, the self-imaging effect for light beams in periodically curved waveguide arrays was observed in AlGaAs waveguides [7-11]. Adiabatic transfer of light in curved waveguide arrays akin to multilevel population transfer of atoms driven by a pulse sequence was discussed in Ref. [12]. Longitudinal modulation of transversally periodic guiding structures allows managing their diffraction properties [13,14] and may even make possible diffractionless propagation of linear beams [15,16]. In the nonlinear regime, the longitudinal modulation of periodic structures gives rise to a variety of parametric phenomena [17,18]. Longitudinal modulations of linear waveguides may stimulate revivals, i.e. periodic transitions between guided modes. The simplest theory of mode transitions in periodically perturbed linear waveguides, leading to a system of coupled differential equations for the mode amplitudes, is well-known today (see, e.g., [19]). Such linear mode conversion can be implemented in photoinduced, long-periodic fiber gratings [20]. However, stimulated mode transitions have not been analyzed for full models that take into account leaking radiation and nonlinearity. Moreover, stimulated mode transitions have been only considered in settings where waveguiding is caused by total internal reflection but not due to more complicated mechanisms, such as, e.g., grating-mediated waveguiding.

In this Letter we predict that Rabi-like flopping and stimulated mode transitions occur with linear and soliton states in properly modulated waveguides. We address a model that takes into account radiation and exact waveguiding mechanism, as well as the solitonic regimes. We start our analysis with linear waveguiding structures and show that suitable longitudinal modulations induce periodic transitions between guided modes of the same parity for general guiding mechanisms. We reveal the possibility of cascade stimulated transitions in systems supporting up to three modes of the same parity. We show that the phenomenon occurs also in the nonlinear regime and in multimode systems, where the mode flopping might be viewed as analogous to interband transitions.

For the sake of generality, we consider the propagation of a laser beam along the $\xi$ axis in a medium with an instantaneous cubic nonlinearity and transverse shallow modulation of refractive index described by the nonlinear Schrödinger equation for dimensionless field amplitude $q$:

$$i\frac{\partial q}{\partial \xi} = -\frac{1}{2}\frac{\partial^2 q}{\partial \eta^2} - pR(\eta)[1 + \mu\cos(\Omega_\xi \xi)]q - \sigma|q|^2 q. \qquad (1)$$

Here the longitudinal $\xi$ and transverse $\eta$ coordinates are scaled to the diffraction length and input beam width, respectively; the function $R(\eta)$ describes the transverse profile of the refractive index; the parameter $p$ is proportional to the depth of the refractive index modulation; the depth of longitudinal periodic modulation with the spatial frequency $\Omega_\xi$ is characterized by the parameter $\mu < 1$; and, finally, the parameter $\sigma = 0$ corresponds to the linear regime while $\sigma = 1$ corresponds to focusing nonlinearity. Technological fabrication of longitudinally-modulated refractive index landscapes is well established [9,10], while optical induction in suitable crystals offers a potential powerful alternative [21].

We start our analysis from the simplest case of a waveguide with Gaussian transverse refractive index profile $R(\eta) = \exp[-(\eta/W)^2]$. Such waveguide supports only three linear modes at $p = 2.3$ and $W = 2$. The stationary mode profiles $w_n(\eta)$ and corresponding propagation constants $b_n$ at $\mu = 0$ are found numerically from the linear eigenvalue problem $bw = (1/2)d^2w/d\eta^2 + R(\eta)w$ that is obtained from Eq. (1) upon substitution $q(\eta, \xi) = w(\eta)\exp(ib\xi)$. For illustrative purposes we depict in Fig. 1(a) the refractive index profile as a potential well, while the profiles of the symmetric modes are depicted as eigenfunctions of the corresponding energy levels. We now set the mode $w_3(\eta)$ as the initial condition for the linear version of Eq. (1) in the presence of a longitudinal modulation of the refractive index $(\mu \neq 0)$.

On physical grounds, one of the central findings of this Letter is that the application of a suitable shallow longitudinal refractive index modulation with the frequency $\Omega_\xi \approx b_1 - b_3$ (here $b_{1,3}$ are the propagation constants of eigenmodes with the same symmetry) stimulates the mode *conversion* $w_3(\eta) \to w_1(\eta)$, in a manner similar to that in Rabi floppy between quantum atomic states. This process is *reversible*, so that the initial eigenmode is recovered after a certain propagation distance, but recovery is not complete because of the small radiation visible in Fig. 1(b). Figures 1(c) and 1(d) illustrate the oscillatory behavior of the field amplitude $|q(0,\xi)|$ in the waveguide center. For shallow longitudinal modulations $(\mu \ll 1)$, the oscillations of $|q(0,\xi)|$ correspond to the superposition of two eigenmodes $q(\eta,\xi) = C_1(\xi)w_1(\eta)\exp(ib_1\xi) + C_3(\xi)w_3(\eta)\exp(ib_3\xi)$, where $C_{1,3}(\xi)$ are complex weight coefficients (thus for the transition $w_3(\eta) \to w_1(\eta)$ one has $C_3(0) = 1$ and $C_1(0) = 0$). The standard technique of analysis of resonant $(\Omega_\xi = b_1 - b_3)$ mode coupling (see Ref. [22]) yields the system of equations

$$i\langle w_1 w_1 \rangle \frac{dC_1}{d\xi} = \frac{1}{2} p\mu \langle w_1 R w_3 \rangle C_3,$$
$$i\langle w_3 w_3 \rangle \frac{dC_3}{d\xi} = \frac{1}{2} p\mu \langle w_3 R w_1 \rangle C_1,$$
(2)

where angular brackets stand for spatial averaging, i.e. $\langle w_1 R w_3 \rangle = \int_{-\infty}^{\infty} w_1(\eta) R(\eta) w_3(\eta) d\eta$. Notice that energy exchange is possible only between modes with *equal parity*. Otherwise the overlap integral $\langle w_i R w_k \rangle$ vanishes for symmetric refractive index distribution $R(\eta)$. Equations (2) can be used to derive the flopping frequency $\Omega_R = (\mu p / 2) \langle w_1 R w_3 \rangle / \langle w_1 w_1 \rangle^{1/2} \langle w_3 w_3 \rangle^{1/2}$. This frequency is proportional to the depth of the longitudinal modulation. Stimulated conversion of modes with equal parities is also possible in systems with periodic variation of waveguide width, but in this setting the radiative losses become more pronounced. Notice that stimulated transitions between eigenmodes with *different* parity may be possible in periodically curved structures, such as those studied in Refs. [8-11], or in waveguides with tilted gratings [20]. Still, in such settings the radiative losses accompanying stimulated transitions are also essential.

The conversion is most pronounced and radiative losses are minimal at *resonance* $\Omega_\xi = b_1 - b_3$. When the longitudinal modulation frequency is slightly detuned from the resonance, the mode conversion is incomplete (the weight coefficient decreases but it does not vanish at $\xi = \xi_{tr}$) and some residual oscillations of the central point amplitude $q_c$ remain in the vicinity of $\xi_{tr}$ (Fig. 1(c)). The contrast of the residual oscillations $V = (q_c^{max} - q_c^{min}) / (q_c^{max} + q_c^{min})$ may be used to quantify the quality of the stimulated mode conversion. Figure 1(e) shows the dependence of the contrast on the relative frequency detuning $\nu$ from resonant value (in percents). As expected on physical grounds, the high-quality mode conversion is possible only in a very narrow band of longitudinal modulation frequencies, which implies that *highly selective* conversion is possible in the presence of multiple eigenmodes. Figure 1(f) shows the dependence of the transition distance $\xi_{tr}$ on the longitudinal modulation depth $\mu$; the decay of the transition distance $\xi_{tr} = \pi / \Omega_R$ with $\mu$ is consistent with the above analytical expression $\Omega_R \sim \mu$. Notice that in the transition point the mode profile corresponding to $b_3$ exactly transforms into the profile corresponding to $b_1$ depicted in Fig. 1(a).

*Highly-selective* stimulated transformation are illustrated in Fig. 2 where we consider five-mode Gaussian waveguide at $W = 2$, $p = 6$. The fifth symmetric mode $w_5(\eta)$ (Fig. 2(a)) is launched into a waveguide with a modulation with $\Omega_\xi = b_5 - b_3$ leading to stimulated conversion $w_5(\eta) \to w_3(\eta)$ at the distance $\xi_{tr}^{(1)} \approx 32$ (Fig. 2(b)). Beyond this point, the frequency of longitudinal modulation was set to $\Omega_\xi = b_3 - b_1$ in order to stimulate transition to the fundamental mode $w_3(\eta) \to w_1(\eta)$ that occurs at $\xi_{tr}^{(2)} \approx 87$. Longitudinal modulation was then suppressed ($\mu \to 0$) beyond this point to avoid reverse conversion. Low-contrast residual oscillations of the amplitude in the center of waveguide for $\xi > \xi_{tr}^{(2)}$ confirm high quality cascaded conversion. The corresponding dynamics is shown in Fig. 2(c). As it was mentioned, in the transition points the mode profiles match those depicted in Fig. 2(a). Note that direct conversion $w_5(\eta) \to w_1(\eta)$ can also be realized at the modulation frequency $\Omega_\xi = b_5 - b_1$, but one finds that the process is accompanied by significantly higher radiative losses.

Another important finding is the persistence of the stimulated mode conversion in the *nonlinear regime*. With nonlinearity the propagation constants, as well as the corresponding mode profiles become functions of the energy flow $U = \int_{-\infty}^{\infty} |q|^2 d\eta$. Figure 3(a) illustrates the shift of the propagation constants as well as change of the transition frequency $b_3(U) - b_1(U)$ with increase of the energy flow $U$ of the modes supported by the nonlinear three-mode Gaussian waveguide ($\sigma = 1$, $W = 2$, $p = 2.3$). This dispersion diagram was calculated by a relaxation method for the eigenvalue problem $bw = (1/2) d^2w / d\eta^2 + R(\eta)w + \sigma w^3$ with $\sigma \neq 0$ obtained from Eq. (1) at $\mu = 0$. The third mode was then selected as initial condition for Eq. (1) with $\mu > 0$. By tuning the frequency of the longitudinal refractive index modulation, it is possible to perform high-quality conversion of nonlinear modes, even for non-negligible values of the energy flow $U \leq 3$. Figures 3(b) and 3(c) show dynamics of such high-quality stimulated nonlinear mode conversion $w_3(\eta) \to w_1(\eta)$ at $U = 1$. The initial and final guided nonlinear states are depicted by circles at dispersion diagram 3(a) for different values of input energy flow, while arrows show the conversion. Increasing $U$ results in higher radiative losses.

Interestingly, the frequency corresponding to the most efficient conversion increases with $U$ too, and it is still approximately given by the value $b_3(U) - b_1(U)$.

The possibility of stimulated conversion of guided nonlinear states is not sensitive to the waveguiding mechanism. Next we show an illustrative example related to grating-mediated guiding [23]. In this case the refractive index profile corresponds to a periodic grating with spatial frequency $\Omega_\eta$ and bell-shaped amplitude modulation, e.g., the Gaussian one: $R(\eta) = (1/2)[1 + \cos(\Omega_\eta \eta)] \exp[-(\eta/W)^2]$. The corresponding refractive index profile (depicted as the potential well) and nonlinear modes with $U = 1$ are shown in Fig. 4(a). One can clearly see that such waveguides also allow high-quality stimulated transitions as well as their simplest Gaussian counterparts despite different type of the mechanism supporting localized modes (Fig. 4(b)).

We also considered the nonlinear analog of interband Rabi flopping which can be experimentally realized in a finite waveguide arrays. The refractive index profile matches an harmonic lattice with the amplitude modulated by the wide super-Gaussian function $R(\eta) = (1/2)[1 + \cos(\Omega_\eta \eta)] \exp[-(\eta/W)^8]$. Such wide waveguide with $\Omega_\eta = 4$, $p = 10$, $W = 8\pi$ supports 44 guided modes whose propagation constants $b_k$ form analogs of "valence" and "conducting" bands separated by the gap (discrete eigenvalues forming those bands are tightly packed and are not resolved in Fig. 4(c) showing eigenvalue spectrum). Figure 4(d) illustrates stimulated conversion of the nonlinear mode from the bottom of "conducting band" into nonlinear mode from the top of the "valence band" in such highly-multimode system (the transition is shown by an arrow in Fig. 4(c) at $U = 1$).

Summarizing, we introduced an analog of Rabi flopping and stimulated transitions in multimode guiding structures. The analogy is found to take place with low-power modes, and also in the presence of weak nonlinearities. Our results motivate new opportunities to efficient state conversion in suitable guiding structures in several physical settings and geometries. We also note that results may be extended to system exhibiting strong longitudinal modulations, by using finite-difference-frequency-domain methods.

# References

1. I. I. Rabi, Phys. Rev. **49**, 324 (1936).
2. R. W. Robinett, Phys. Rep. **392**, 1 (2004).
3. S. Minardi *et al.*, Opt. Lett. **27**, 2097 (2002).
4. L. Deng *et al.*, Phys. Rev. Lett. **83**, 5407 (1999).
5. Y. B. Ovchinnikov, Opt. Commun. **182**, 35 (2000).
6. R. Iwanow *et al.*, Phys. Rev. Lett. **95**, 053902 (2005).
7. G. Lenz *et al.*, Opt. Commun. **218**, 87 (2003).
8. S. Longhi, Opt. Lett. **30**, 2137 (2005).
9. S. Longhi *et al.*, Phys. Rev. Lett. **96**, 243901 (2006).
10. R. Iyer *et al.*, Optics Express **15**, 3212 (2007).
11. I. L. Garanovich, A. A. Sukhorukov, and Y. S. Kivshar, Phys. Rev. E **74**, 066609 (2006).
12. S. Longhi, Phys. Lett. A **359**, 166 (2006).
13. H. S. Eisenberg *et al.*, Phys. Rev. Lett. **85**, 1863 (2000).
14. M. J. Ablowitz and Z. H. Musslimani, Phys. Rev. Lett. **87**, 254102 (2001); T. Pertsch, U. Peschel, and F. Lederer, Chaos **13**, 744 (2003).
15. H. Kosaka *et al.*, Appl. Phys. Lett. **74**, 1212 (1999); D. N. Chigrin *et al.*, Opt. Express **11**, 1203 (2003); R. Illiew *et al.*, Appl. Phys. Lett. **85**, 5854 (2004); D. W. Prather *et al.*, Opt. Lett. **29**, 50 (2004); P. T. Rakich *et al.*, Nat. Mater. **5**, 93 (2006).
16. K. Staliunas and R. Herrero, Phys. Rev. E **73**, 016601 (2006); K. Staliunas, R. Herrero, and G. J. de Valcarcel, Phys. Rev. A **75**, 011604(R) (2007).
17. Y. V. Kartashov, L. Torner, and V. A. Vysloukh, Opt. Lett. **29**, 1102 (2004); Y. V. Kartashov, V. A. Vysloukh, and L. Torner, Phys. Rev. E **71**, 036621 (2005).
18. Y. V. Kartashov, L. Torner, and D. N. Christodoulides, Opt. Lett. **30**, 1378 (2005); C. R. Rosberg *et al.*, Opt. Lett. **31**, 1498 (2006).
19. D. Marcuse, *Theory of Dielectric Optical Waveguides* (Academic, San Diego, Calif. 1991).

20. K. Hill *et al.*, Electron. Lett. **26**, 1270 (1990); K. S. Lee and T. Erdogan, Appl. Optics **39**, 1394 (2000).

21. J. W. Fleischer *et al.*, Nature **422**, 147 (2003).

22. H. A. Haus, *Waves and Fields in Optoelectronics* (Englewood Cliffs, NJ. Prentice-Hall, 1984).

23. O. Cohen *et al.*, Phys. Rev. Lett. **93**, 103902 (2004).

# Figure captions

Figure 1 (color online). (a) Linear modes of three-mode Gaussian waveguide. Black line shows potential well, while each mode profile is shifted in vertical direction to show the position of corresponding energy levels inside the well. Arrows show transition directions. (b) Revival dynamics at $\Omega_\xi = 0.95(b_1 - b_3)$, $\mu = 0.15$ for input mode $w_3(\eta)$. Amplitude $q_c$ versus $\xi$ for resonant transitions $w_3(\eta) \to w_1(\eta)$ (c) and $w_1(\eta) \to w_3(\eta)$ (d) at $\Omega_\xi = b_1 - b_3$, $\mu = 0.15$. (e) Contrast versus $\nu$ for transition $w_3(\eta) \to w_1(\eta)$ at $\mu = 0.15$. (f) Distance of mode conversion $w_3(\eta) \to w_1(\eta)$ versus $\mu$ at $\Omega_\xi = b_1 - b_3$. In all cases $W = 2$, $p = 2.3$.

Figure 2 (color online). (a) Linear modes of five-mode Gaussian waveguide. (b) Amplitude $q_c$ versus $\xi$ for cascade conversion $w_5 \to w_3 \to w_1$ at $\mu = 0.1$. Dashed lines indicate distances where modulation frequency changes from $\Omega_\xi = b_3 - b_5$ to $\Omega_\xi = b_1 - b_3$ and where modulation is switched off. (c) Dynamics of transition $w_5 \to w_3 \to w_1$ corresponding to (b). In all cases $W = 2$, $p = 6$.

Figure 3 (color online). (a) Dispersion diagrams for first and third modes of nonlinear Gaussian waveguide. Arrows show resonant nonlinear transitions $w_3 \to w_1$ at $\mu = 0.15$, where initial and final states are marked by circles. (b) Amplitude $q_c$ versus $\xi$ for resonant transition at $U = 1$, $\Omega_\xi = 1.603$, $\mu = 0.15$. Dashed line indicates the distance where modulation is switched off. (c) Dynamics of transition $w_3 \to w_1$ corresponding to (b). In all cases $W = 2$, $p = 2.3$.

Figure 4 (color online). (a) Linear modes of three-mode grating waveguide at $\Omega = 4$, $W = \pi$, $p = 2$. (b) Dynamics of nonlinear transition $w_3 \to w_1$ in such waveguide at $U=1$, $\mu=0.15$, $\Omega_\xi = 0.662$. Longitudinal modulation is switched off at $\xi = \xi_{\rm tr} = 122.8$. (c) Band structure of propagation constants in super-Gaussian modulated waveguide with $\Omega = 4$, $W = 8\pi$, $p = 10$ (tightly packed discrete eigenvalues, assembled into bands, are not resolved). (d) Dynamics of transition between the bottom of "conductivity band" and the top of "valence band" at $U = 1$, $\mu = 0.15$, $\Omega_\xi = 4.92$.

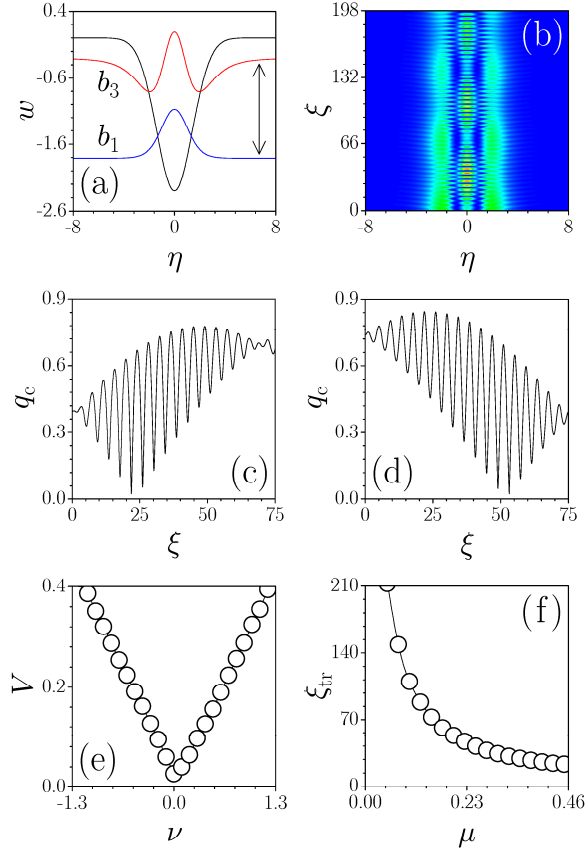

Figure 1 (color online). (a) Linear modes of three-mode Gaussian waveguide. Black line shows potential well, while each mode profile is shifted in vertical direction to show the position of corresponding energy levels inside the well. Arrows show transition directions. (b) Revival dynamics at $\Omega_\xi = 0.95(b_1 - b_3)$, $\mu = 0.15$ for input mode $w_3(\eta)$. Amplitude $q_c$ versus $\xi$ for resonant transitions $w_3(\eta) \to w_1(\eta)$ (c) and $w_1(\eta) \to w_3(\eta)$ (d) at $\Omega_\xi = b_1 - b_3$, $\mu = 0.15$. (e) Contrast versus $\nu$ for transition $w_3(\eta) \to w_1(\eta)$ at $\mu = 0.15$. (f) Distance of mode conversion $w_3(\eta) \to w_1(\eta)$ versus $\mu$ at $\Omega_\xi = b_1 - b_3$. In all cases $W = 2$, $p = 2.3$.

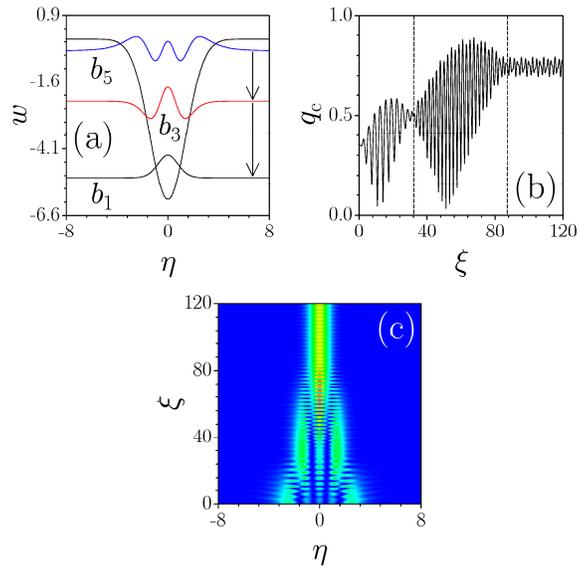

Figure 2 (color online). (a) Linear modes of five-mode Gaussian waveguide. (b) Amplitude $q_c$ versus $\xi$ for cascade conversion $w_5 \to w_3 \to w_1$ at $\mu = 0.1$. Dashed lines indicate distances where modulation frequency changes from $\Omega_\xi = b_3 - b_5$ to $\Omega_\xi = b_1 - b_3$ and where modulation is switched off. (c) Dynamics of transition $w_5 \to w_3 \to w_1$ corresponding to (b). In all cases $W = 2$, $p = 6$.

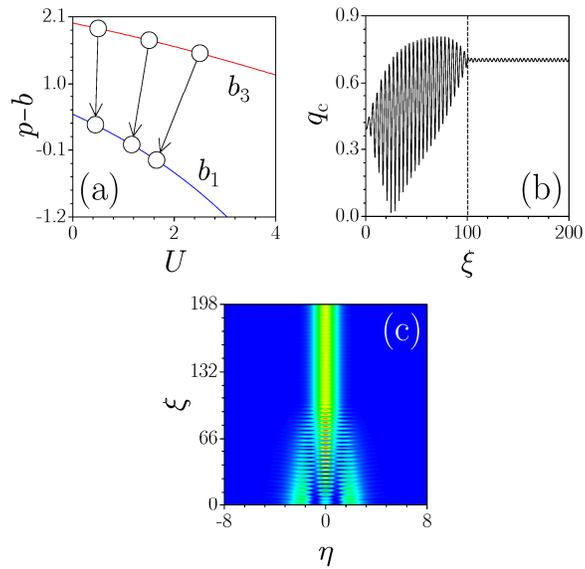

Figure 3 (color online). (a) Dispersion diagrams for first and third modes of nonlinear Gaussian waveguide. Arrows show resonant nonlinear transitions $w_3 \to w_1$ at $\mu = 0.15$, where initial and final states are marked by circles. (b) Amplitude $q_c$ versus $\xi$ for resonant transition at $U = 1$, $\Omega_\xi = 1.603$, $\mu = 0.15$. Dashed line indicates the distance where modulation is switched off. (c) Dynamics of transition $w_3 \to w_1$ corresponding to (b). In all cases $W = 2$, $p = 2.3$.

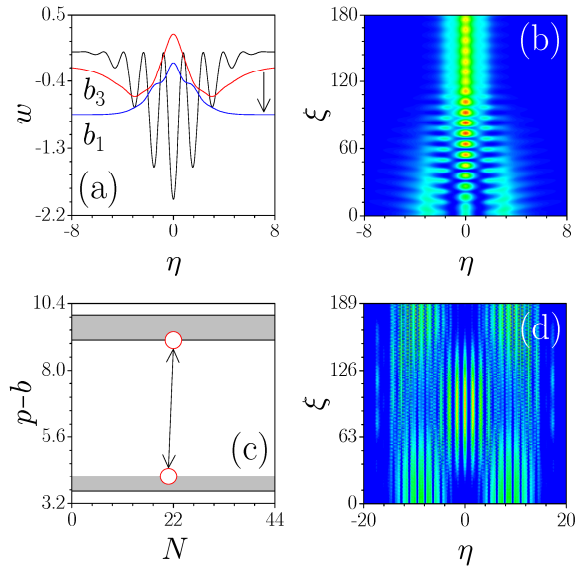

Figure 4 (color online). (a) Linear modes of three-mode grating waveguide at $\Omega=4$, $W=\pi$, $p=2$. (b) Dynamics of nonlinear transition $w_3 \to w_1$ in such waveguide at $U=1$, $\mu=0.15$, $\Omega_\xi=0.662$. Longitudinal modulation is switched off at $\xi=\xi_{\rm tr}=122.8$. (c) Band structure of propagation constants in super-Gaussian modulated waveguide with $\Omega=4$, $W=8\pi$, $p=10$ (tightly packed discrete eigenvalues, assembled into bands, are not resolved). (d) Dynamics of transition between the bottom of "conductivity band" and the top of "valence band" at $U=1$, $\mu=0.15$, $\Omega_\xi=4.92$.

14